\documentclass[a4paper,fleqn,usenatbib]{mnras}

\usepackage[T1]{fontenc}
\usepackage{ae,aecompl}

\usepackage{graphicx}	
\usepackage{amsmath}	
\usepackage{amssymb}	
\usepackage[lofdepth,lotdepth]{subfig}

\title[Variable Near-Infrared AGN in UKIDSS UDS]{Long-term NIR Variability in the UKIDSS Ultra Deep Survey: a new probe of AGN activity at high redshift.}

\author[E. Elmer et al.]{E. Elmer,$^{1}$\thanks{E-mail: ppxee@nottingham.ac.uk}
O. Almaini,$^{1}$
M. Merrifield,$^{1}$
W. G. Hartley,$^{2}$
D. T. Maltby, $^{1}$
\newauthor A. Lawrence,$^{3}$
I. Botti,$^{4}$
and P. Hirst$^{5}$
\\
$^{1}$School of Physics and Astronomy, University of Nottingham, University Park, Nottingham, NG7 2RD, UK\\
$^{2}$Department of Astronomy, University of Geneva, CH-1205 Versoix, Switzerland\\
$^{3}$Institute for Astronomy, University of Edinburgh, Royal Observatory, Edinburgh, EH9 3HJ, United Kingdom\\
$^{4}$Facultad de Ingenier\'ia, Universidad del Desarrollo. Av. Plaza 680, Las Condes, Santiago, Chile\\
$^{5}$Gemini Observatory, 670 N. Aohoku Place, Hilo, HI 96720, USA.\\
}

\date{Accepted XXX. Received YYY; in original form ZZZ}

\pubyear{2019}

\begin{document}
\label{firstpage}
\pagerange{\pageref{firstpage}--\pageref{lastpage}}
\maketitle

\begin{abstract}
    We present the first attempt to select AGN using long-term NIR variability. By analysing the $K$-band light curves of all the galaxies in the UKIDSS Ultra Deep Survey, the deepest NIR survey over $\sim 1$ sq degree, we have isolated 393 variable AGN candidates. A comparison to other selection techniques shows that only half of the variable sources are also selected using either deep Chandra X-ray imaging or IRAC colour selection, suggesting that using NIR variability can locate AGN that are missed by more standard selection techniques. In particular, we find that long-term NIR variability identifies AGN at low luminosities and in host galaxies with low stellar masses, many of which appear relatively X-ray quiet.
    
\end{abstract}




\section{Introduction}
\label{sec:intro}

    Active Galactic Nuclei (AGN) are some of the most luminous objects in the known Universe, believed to be powered by accretion onto supermassive black holes \citep[e.g.][]{Rees1984}.
		
	There are a variety of methods that can be used to select AGN from pre-existing catalogues, using both photometric \citep[e.g.][]{Stern2005, Lacy2004, Lacy2007} and spectroscopic \citep[e.g.][]{Feltre2016} criteria, and each of them has both strengths and weaknesses. For example, some methods are geared towards low redshift samples, and can become highly contaminated at high redshifts \citep{Donley2008}. \cite{Review2017} gives an overview of identification techniques across the electromagnetic spectrum, and their selection effects. Selecting AGN through X-ray emission is generally accepted to be the most complete method, but this will only find the brightest AGN at high redshifts.
	
	It is well established that the light from AGN changes over time \citep[e.g.][]{Angione1973}. This characteristic can be used to select AGN, and to study the physics of the system on scales that are difficult to resolve with standard observational techniques. Selection by variability has been previously undertaken in the optical/UV \citep[e.g.][]{Pouliasis2019, Sanchez-Saez2019} and X-ray \citep[e.g.][]{Young2012} regimes. 
	
	The properties of variability have been studied in many wavebands \citep[e.g.][in optical, X-ray, and NIR respectively]{Angione1973, Marshall1981, Neugebauer1989} but there is not yet a standard, conclusive model for what causes the observed variations. The timescale for variability is known to increase as the detection wavelength increases. In X-rays, variability is seen on scales of minutes to hours, while in optical it is on scales of days to months, and in the near-infrared (NIR) significant variability is only seen over months or years. Because of these changes in timescales, the current theory postulates that the variability originates from stochastic variations in the X-ray corona. These then propagate out through the AGN, creating the variations in emission seen from the accretion disk (optical) and the torus (IR) \citep[e.g.][]{McHardy2016,Clavel1992}. However, this theory has been called into question by evidence of intrinsic optical and UV variability over long timescales \citep[e.g.][]{Arevalo2008,Arevalo2009}.
	
	Due to the long timescales, NIR variability is largely unexplored at high redshifts. This band is particularly interesting, however, as it probes rest-frame red optical light at high redshifts, allowing comparisons to varibility in the local universe. In addition, rest-frame NIR light has been shown to originate in both the accretion disk and the dusty torus \citep{Lira2011, Landt2011}. Previous NIR studies have generally either used observations of previously known AGN to confirm that AGN do vary in the NIR \citep[e.g.][]{Enya2002a} or used ensemble variability to characterise the variations \citep[e.g.][]{Kouzuma2011}. 
	
	Some of the longest NIR AGN light curves studied to date are presented in \cite{Neugebauer1989} and \cite{Sanchez2017}. \cite{Neugebauer1989} explored the NIR variability of individual AGN by using a number of different datasets to construct light curves spanning an average of 6 years, with some covering up to 20 years. 
    \cite{Sanchez2017} examined the NIR light curves of individual AGN over approximately 5 years, using photometric data from the UltraVISTA survey \citep{McCracken2012}. 
    In both cases, the AGN were selected in different wavebands: \cite{Neugebauer1989} used optical selection, while \cite{Sanchez2017} studied a sample of X-ray selected AGN.
	
	Selection by NIR variability has not previously been attempted. This is because a survey with a long baseline of observations, approximately 5 to 10 years, is required to detect the high amplitude variations. In addition to this, deep data is required to allow the detection of faint, high redshift objects and studies of small amplitude variability, and a wide survey area is preferred in order to find rare types of AGN from just the variability data.
	The majority of variations in the NIR light curves of AGN are only expected to be observable from year to year. Although there is evidence of small amplitude variations on shorter timescales \citep[e.g.][]{Sanchez2017}, AGN variability typically has a red noise power spectrum so the largest amplitude variations are seen over long timescales \citep[e.g.][]{Lawrence1987}. As such, selection through NIR variability on long timescales ought to provide a relatively clean sample of AGN; the main contaminants are expected to be variable stars and supernovae.
	
	The $\sim 8$ year baseline of the United Kingdom Infrared Telescope (UKIRT) Infrared Deep Sky Survey (UKIDSS) Ultra Deep Survey (UDS) makes it the perfect data set for a study of this kind. The long baseline probes the timescales where previous works have found interesting variability, and, as the deepest NIR survey over such a wide area ($5\sigma$ detection limit of $K=25.3$ AB over $\sim 1\ \mathrm{deg^{2}}$), it is deep enough to find faint AGN, while still covering a large enough area to find rare objects amongst the AGN population.
	
	This work focuses on selecting AGN using only their variability in the NIR, and we compare the sample selected in this way to AGN selected using X-ray emission, and infrared (IR) colours. 
	The unique nature of the UDS, combined with removing any prejudice introduced by prior selection techniques, means that this study can probe AGN in ordinary host galaxies as well as bright quasars.

    The selection method is presented in Section \ref{sec:method}. We then compare to other selection techniques in Section \ref{sec:comparsion}, and present a population of X-ray quiet, low stellar mass, low-luminosity AGN found through the analysis in Section \ref{sec:newpop}. We adopt a $\Lambda$CDM cosmology with $H_{0} = 70 \mathrm{\ km s^-1 \ Mpc^{-1}}, \ \Omega_{\Lambda}=0.7,\ \Omega_{m}=0.3$.

\section{Method}
    \label{sec:method}

    \subsection{Data}
        \label{sec:data} 
    
        The UDS provides a unique data set for studying NIR variability of AGN. In addition to being the deepest NIR survey over $1\ \mathrm{deg^{2}}$, with a $5\sigma$ detection limit of $K=25.3$ (AB), the UDS has an unprecedented 8 year baseline for time domain studies. The field is also well studied in a wide range of wavebands, including X-ray imaging of the whole field to a depth of $\sim 10^{-15}\ \mathrm{erg\ cm^{-2}\ s^{-1}}$ with XMM \citep{Ueda2008}, deep Chandra imaging of the centre of the field \citep[$F_{lim} \sim 10^{-16}\ \mathrm{erg\ cm^{-2}\ s^{-1}}$][]{Kocevski2018}, and imaging in all four bands of the Spitzer Infrared Array Camera \citep[IRAC,][]{Fazio2004,Dunlop2007}.
        
        Photometric redshifts were determined using the deep 12-band photometry available in the UDS field ($U,B,V,R,i^{\prime} ,z^{\prime} ,Y,J,H,K,3.6,4.5$), based on the method outlined in \cite{Simpson2013}. A wide range of galaxy templates were used, using simple stellar populations from \cite{Bruzual2003}, with fitting completed using the EAZY software \citep{Brammer2008}. The resulting photometric redshifts were then used to derive other galaxy properties, such as luminosities and stellar masses. Further details will be provided in Almaini et al. (in prep) and Hartley et al. (in prep). 
        
        In addition to deep photometric data, there has been extensive spectroscopic follow-up of the UDS field. The main follow-up programme was the UDSz (ESO large programme 180.A-0776), which used the VIMOS and FORS2 instruments on the ESO VLT to obtain spectra for $>3500$ galaxies \citep[see][]{Bradshaw2013,McLure2013}. Furthermore, the field was covered by the VANDELS spectroscopic survey \citep[ESO programme 194.A-2003,][]{McLure2018,Pentericci2018}, providing $\sim 780$ VIMOS spectra, and a follow up of post-starburst galaxies using VIMOS has also been completed, providing $\sim 100$ additional spectra \citep{Maltby2016}. 
        A further $\sim 4000$ spectroscopic redshifts are provided from archival data, details of which can be found in \cite{Simpson2012} and references therein.
    			
    	The near-infrared imaging in $J$, $H$, $K$ was obtained using the Wide Field Camera \citep[WFCAM;][]{Casali2007} at UKIRT. A single WFCAM observation of the UDS comprises 10s exposures with a 3 x 3 micro-step, in order to increase the pixel resolution of the images, and a 3 x 3 dither, which offsets the effects of bad pixels and flat-fielding complications. This adds up to a total exposure time of 810s per observation. As WFCAM consists of four 13.7 x 13.7 arcmin detectors with 12.9 arcmin between them, a total of at least 4 observations offset by $\sim 13$ arcmin is required to cover the whole field. Further details on the UDS observing strategy can be found in \cite{Lawrence2007} and Almaini et al. (in prep).
    			
    	The images used in this analysis were stacks of all acceptable $K$-band observations taken during one semester of observing. 
    	We chose to complete this study on semester stacks as using the stacked semester images allows us to probe faint, high redshift objects, and preliminary tests revealed that the majority of the variability could be seen on those timescales. This is expected as AGN variability follows a red noise pattern where the largest amplitude variability is seen on longer timescales \citep[e.g.][]{Lawrence1987}. In addition, the image preparation techniques used in this work are non-trivial on shallower stacks, and probing shorter timescales would considerably increase the amount of contamination by supernova and other variables. Shorter timescale stacks will be examined in future work when looking into the structure of the variability. 
    	The $K$-band was chosen for this selection as the data was taken in the best seeing, it has the longest exposure time, and it is the deepest band for the selection of typical high-redshift galaxies. The $J$ and $H$-bands will be added to the analysis in future work, to explore the correlation of variability in different wavebands.
    	
\begin{figure}
			\centering
			\includegraphics[width=\columnwidth]{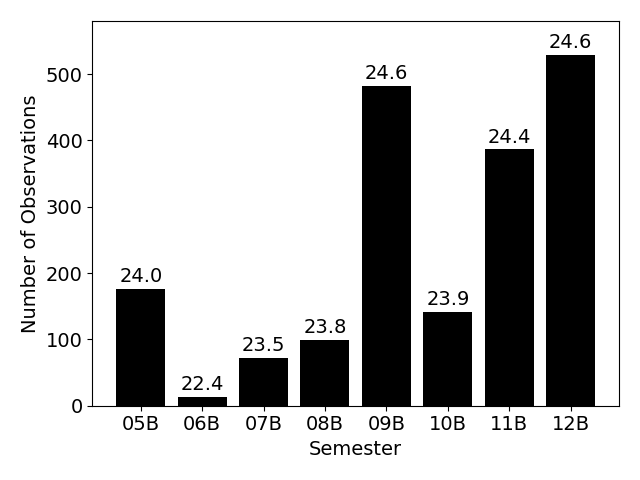}

            \caption{Bar chart to show the number of 810s observations included in each semester stack. The numbers at the top of the bars indicate the $K$-band $5\sigma$ detection limits in AB magnitudes for each of the semester stacks. The 06B semester did not have enough observations to be included in the analysis.}
			\label{fig:semesternumbers}
		\end{figure}

    	As the UDS field is only visible in the second semester of a year, there are a total of 8 epochs across the whole survey. Unfortunately, not enough observations were taken in the 2006B semester (Figure \ref{fig:semesternumbers}), and therefore the image is dominated by noise. As such, 2006B was not included, and the final analysis was undertaken using 7 epochs spanning 2005 to 2012. Photometry was extracted from these stacks by placing 2 arcsec apertures at all the positions where an object was found in the final DR11 image of the UDS using the SExtractor software \citep{Bertin1996}. A 2 arcsec aperture was chosen as it is optimises the trade off between the aperture being large enough to minimise the effect of changing seeing on the light curves, and small enough to minimise dilution of the variability by the host galaxy.
    			
    \subsection{Selection Technique}
        \label{sec:selection} 
        
\begin{figure}
		    \centering
			\includegraphics[width=\columnwidth]{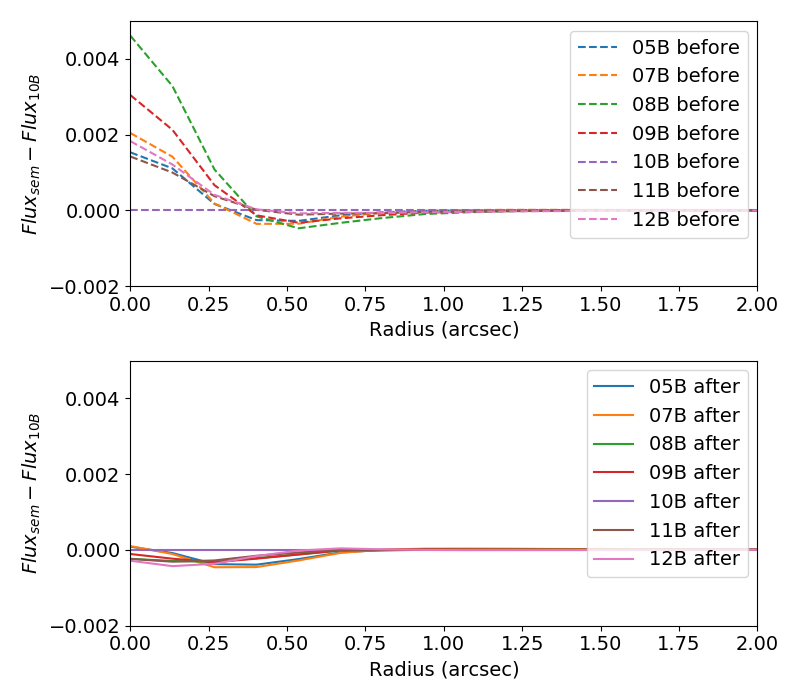}

            \caption{The difference between the radial profiles of the PSFs in each semester relative to 10B, before (top) and after (bottom) convolution. The PSFs were created using $\sim 2000$ stars in the magnitude range $15<K_{vega}<19$. The differences seen in the top panel are drastically reduced by the convolution.}        
			\label{fig:rpcomp}
		\end{figure}

        In order to ensure that the variations seen in the light curves of objects are truly those of the observed object, any other sources of variation must be taken into account. The main source of variations between semester stacks is changes in the point spread function (PSF). In order to minimise this effect, each stack was convolved with a Gaussian kernel to match the PSFs to the semester with the poorest PSF, which was the 2010B stack. The PSFs were constructed by creating an image of the median star based on a sample of $\sim 2000$ stars in the magnitude range $15<K_{vega}<19$.
		The impact of this convolution is best visualised by examining the differences between radial profiles of the PSFs before and after convolution. Figure \ref{fig:rpcomp} shows that before the convolution there are large differences between the radial profiles, especially around the central peak, but these differences are largely eliminated by the convolution.			
			
\begin{figure}
		    \centering
			\includegraphics[width=\columnwidth]{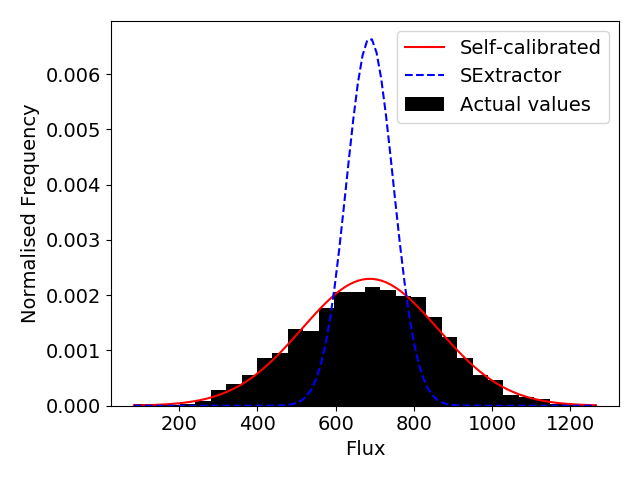}

            \caption{The distribution of flux values within a single quadrant, epoch and flux bin. The red solid line indicates the Gaussian fit to the data when the self-calibrated $\sigma_{F}$ value is used. The blue dashed line shows the Gaussian fit using the median $\sigma_{F}$ value outputted from SExtractor in that bin.}        
			\label{fig:errdemo}
		\end{figure}

\begin{figure}
            \centering
        	\includegraphics[width=\columnwidth]{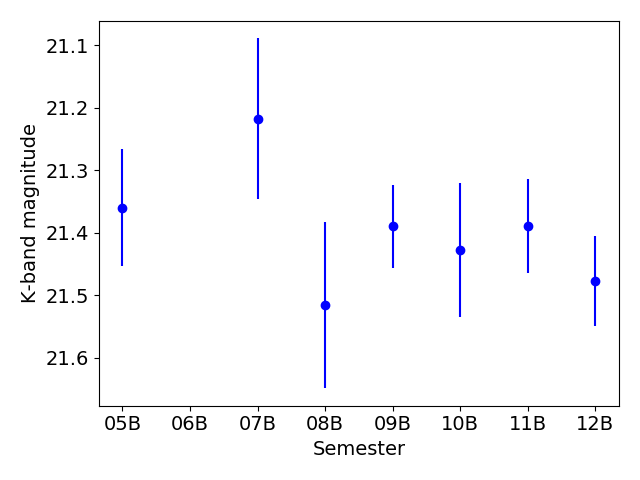}

        	\caption{An example of a non-variable light curve with $\chi^{2} = 4.26$.}
        	\label{fig:lc1}
        \end{figure}

\begin{figure}
            \centering
        	\includegraphics[width=\columnwidth]{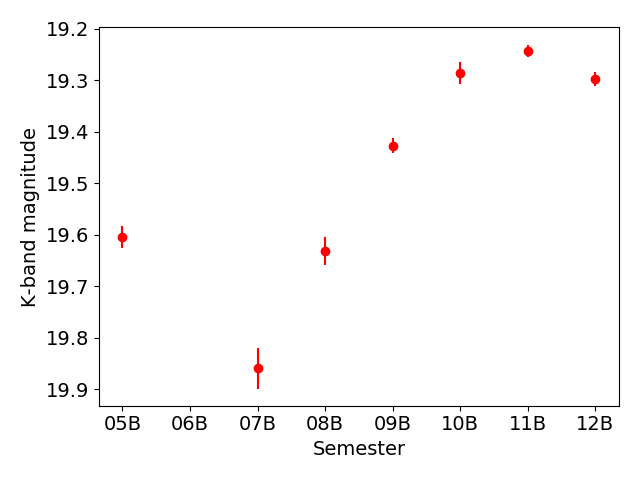}

        	\caption{An example of a variable light curve with $\chi^{2} = 710.03$.}
        	\label{fig:lc2}
        \end{figure}

		As the photometric uncertainties from SExtractor are typically underestimated \citep{Molino2014, Sonnett2013}, we instead use uncertainties that were self-calibrated from the data. The method used was tailored to take into account any remaining variations due to noise or systematic effects.
		In order to ensure these effects were incorporated into the uncertainties, we used the spread of flux values in the non-varying sources to provide a direct estimate of the uncertainty on the flux. The varying sources were excluded from the analysis by modelling the uncertainties as the characteristic variance with flux, and running a preliminary $\chi^{2}$ analysis to remove those with significant variability. This process was then iterated to ensure the model wasn't skewed by the varying sources. The light curves of non-varying sources were then normalised and split into bins of flux (determined from the overall stack), epoch and quadrant. Within each bin, the standard deviation of the normalised flux values then provided the calibrated photometric uncertainty on any object in that bin. This process and the improvement from SExtractor errors is demonstrated in Figure \ref{fig:errdemo}.
		Examples of the resulting light curves with these uncertainties are shown in Figures \ref{fig:lc1} and \ref{fig:lc2}.
			
		Before the selection was completed the catalogue was masked to exclude regions that are too close to bright stars or effected by image artefacts, and any objects that had negative flux values in any semester were removed from the catalogue. We acknowledge that some of the most extreme variables may be removed in this step, but for our purposes we require a visible source in every epoch in order to ensure that our photometric redshifts are as reliable as possible (see Section \ref{sec:comparsion}). These extreme events will be studied in future work. We also required that each object had optimal coverage in all 12 photometric bands. Known galactic stars were also excluded from the analysis as they are the only other objects expected to vary on these timescales. These stars were identified using a combination of SED fitting and image profiling that robustly separates stars and galaxies based on both colours and morphology \citep[for more details see][and Almaini et al., in prep]{Simpson2013}; a total of 5592 were excluded. Supernova are unlikely to be a significant source of contamination on yearly timescales as all but the most extreme cases will average out within the semester stack. The original catalogue contains 296,036 sources; this is reduced to 152,682 sources by these criteria.
		
\begin{figure}
		    \centering
			\includegraphics[width=\columnwidth]{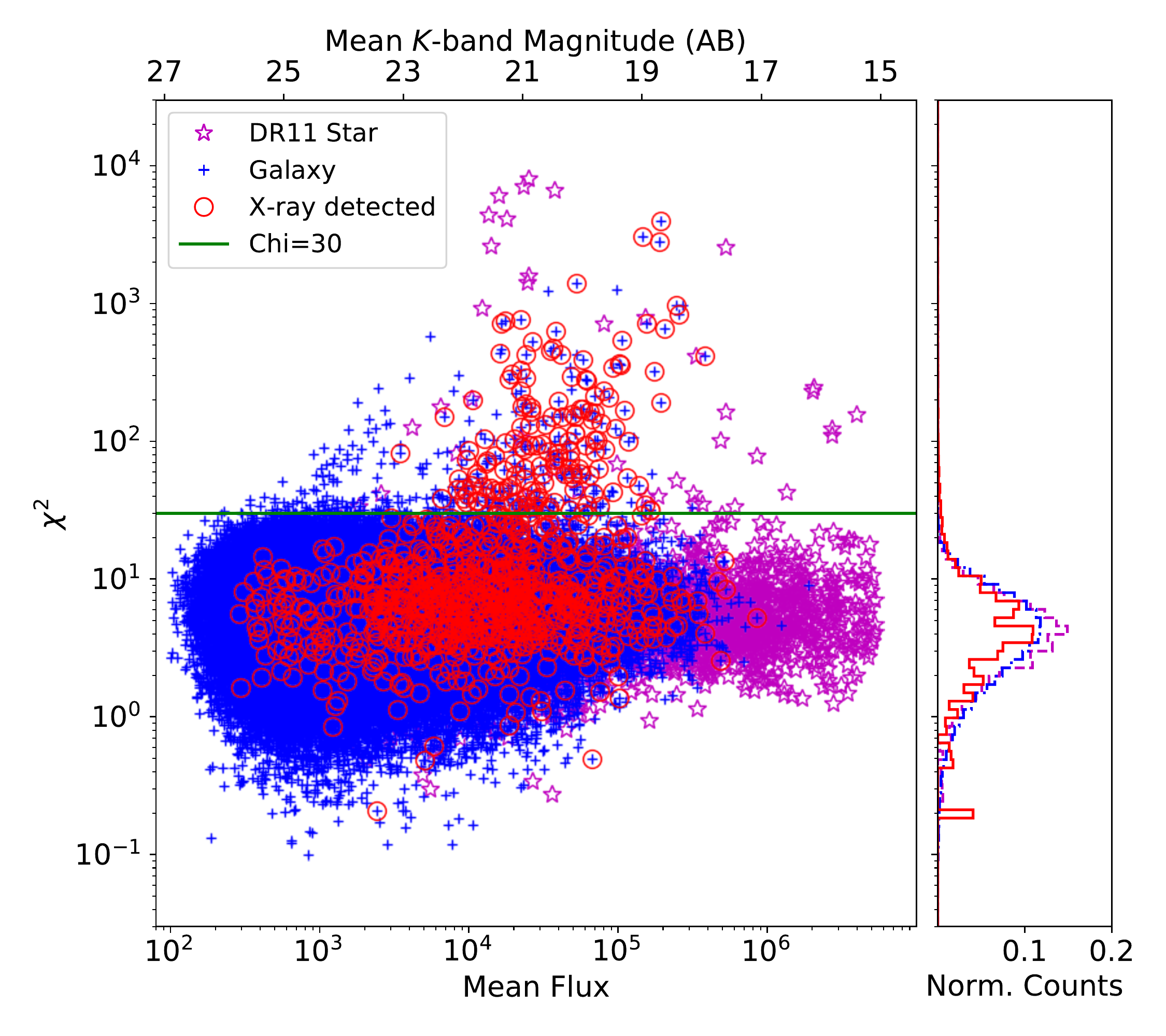}

			\caption{$\chi^{2}$ against mean $K$ band flux for all the objects in the UDS field that are observed in all 7 semesters, the green line indicates the selection threshold.}
			\label{fig:chisquared}
		\end{figure}

		The variable source selection method involves constructing a light curve for each object from the 7 semester stacks, and running a $\chi^{2}$ analysis. For this work we define
		\begin{equation}
		    \centering
		    \chi^{2}=\sum_{i}\frac{(F_{i} - \bar{F})^{2}}{\sigma_{i}^{2}}
		\end{equation}
		where $F_{i}$ is the flux of an object in an epoch, $\bar{F}$ is the average flux of that object across all epochs, and $\sigma_{i}$ is the uncertainty on $F_{i}$. The $\chi^{2}$ statistic provides a simple test of the null hypothesis that a source is non-variable.
		The complete selection is demonstrated in Figure \ref{fig:chisquared}, which shows the distribution of $\chi^{2}$ values as a function of flux. A threshold of $\chi^{2} > 30$ was used to define which sources are considered variable; this value was chosen as it minimises the expected number of false positives while maximising the number of variables found. From a $\chi^{2}$ distribution with 6 degrees of freedom within each flux bin, we expect less than one false positive above this threshold. Over the whole data set, a total of 6 false positives would be expected. When applied to the reduced catalogue described above, the method found a total of 393 variable sources, representing a density of $\mathrm{\sim 550/deg^{2}}$.
		
		Although this sample does not include any objects identified as stars in the UDS catalogue, we completed independent colour and morphology tests to ensure there was minimal stellar contamination. These checks involved quantifying whether the light profiles of the objects appeared 'stellar', and using colour-colour plots \citep[such as those used in][]{Sanchez-Saez2019,Lane2007} to examine whether the variables had colours consistent with stars. All of the 393 variables had galaxy morphologies and optical/IR colours that are not consistent with the stellar locus.
	
\section{Comparison to other AGN populations}
    \label{sec:comparsion}

    \subsection{X-ray AGN}
        \label{sec:xray}
        
        If a high redshift source is bright in X-rays, there as a very high likelihood that it is an AGN. As such, deep X-ray imaging is a clean and relatively complete method of finding AGN in deep sky surveys \citep{Review2017}. The main contaminants in X-ray imaging are stars, but in the UDS the star/galaxy separation is generally reliable, and therefore stars are easily identified and removed. 
			
		The X-ray AGN catalogue was constructed from the deep Chandra imaging covering the central region of the UDS field \citep{Kozowski2016}. Within this region, there are 593 X-ray AGN with associated $K$-band detections in the DR11 catalogue; sources were matched between X-ray and NIR data to within 1 arcsec. 
		74 (41\%) of the 181 variable sources within the Chandra field are in the X-ray AGN sample, which is convincing evidence that the variability method does detect AGN. This proportion remains relatively constant if the analysis is extended to include the shallower XMM imaging of the whole field (167, 42\%, of 393 variables), indicating the non-detections are not biased by the flux limits of the surveys (see Section \ref{sec:xraynonxray}). The relative numbers in each sample are outlined in Table \ref{tab:xraynumbers}.
		
		\begin{table}
		    \begin{center}
		    	\resizebox{\columnwidth}{!}{%
    		    \begin{tabular}{| c | c | c |}
    		        \hline
    		         & Chandra Region & Full Field  \\
    		         & (Chandra)      & (XMM and Chandra) \\
    		        \hline
    		        X-ray Selected AGN (A)        & 593  & 1247 \\
    		        Variability Selected AGN (B)  & 181  & 393 \\
    		        \% of A that are also B       & 12\% & 13\% \\
    		        \% of B that are also A       & 41\% & 42\% \\
    		        \hline
    		    \end{tabular}}
		    \end{center}
		    \caption{Summary of the X-ray and NIR-Variable AGN samples. Numbers in each set are shown for the central region imaged with Chandra and the full field imaged with both XMM and Chandra.}
		    \label{tab:xraynumbers}
		\end{table}
		
\begin{figure*}
			\centering
			\includegraphics[width=\textwidth]{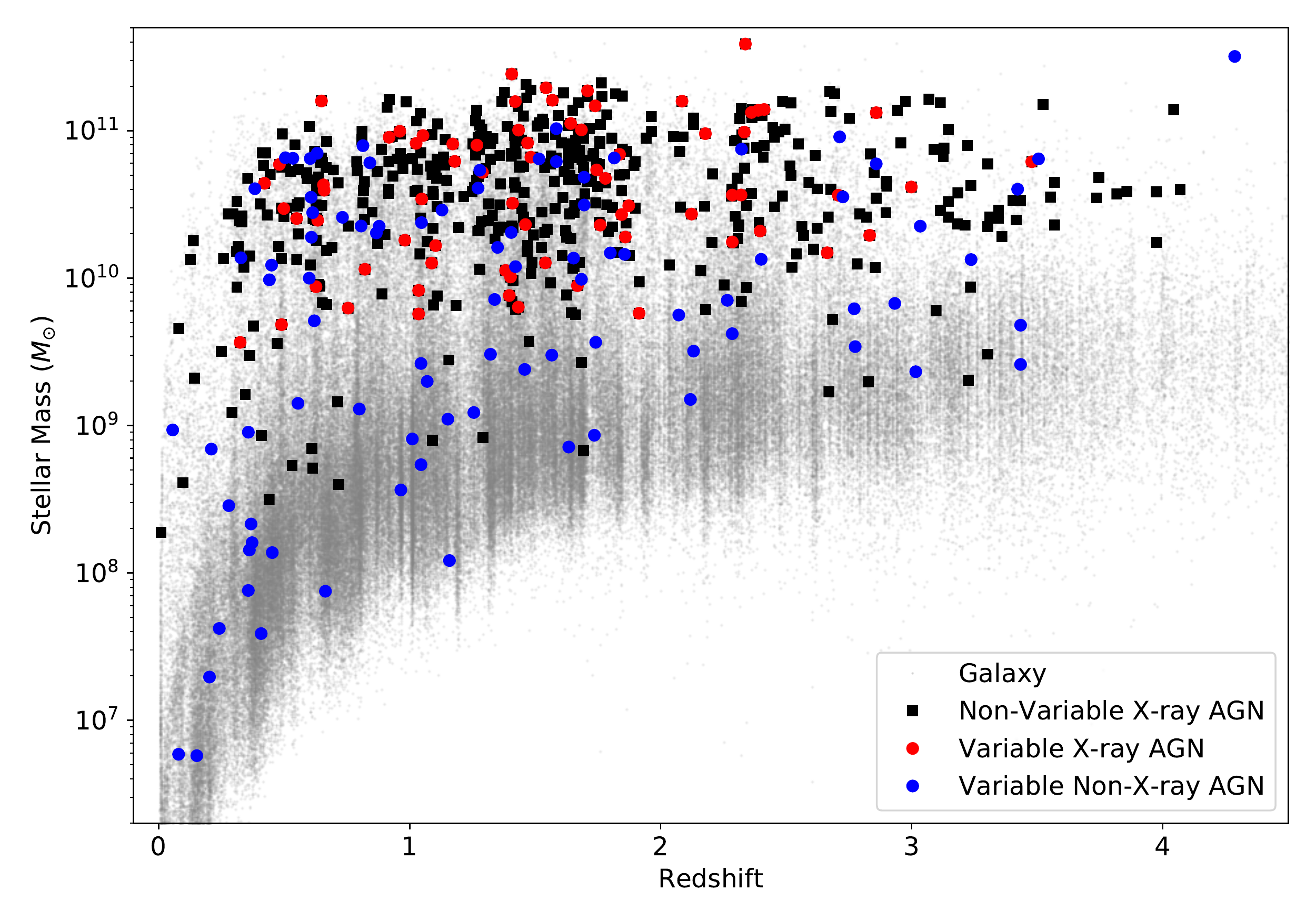}

			\caption{Stellar mass distribution with redshift showing where the X-ray and variable AGN populations sit within the parameter space, restricted to only the sources within the Chandra region. Variability selected AGN are found to occupy a wider range in stellar mass.}
			\label{fig:mstarxray}
		\end{figure*}

		Figure \ref{fig:mstarxray} shows the distribution of stellar masses with redshift for all galaxies in the UDS. When the X-ray and variable AGN populations within the Chandra region are highlighted, it is clear that the variability analysis finds AGN across the parameter space, whereas the deep Chandra imaging only probes the high mass end. A KS test of the stellar mass distributions confirms that the variable and X-ray populations have significantly different distributions in stellar mass, with a p-value of $7.8\times10^{-13}$. 
		
\begin{figure}
			\centering
			\includegraphics[width=\columnwidth]{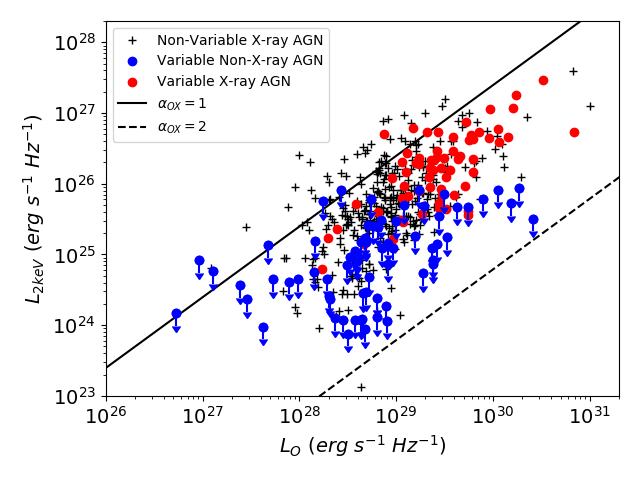}
			\caption{Monochromatic X-ray luminosity at 2 keV against monochromatic optical luminosity at 2500 \AA. The solid line indicates where $\alpha_{OX} = 1$, and the dashed line shows $\alpha_{OX} = 2$, where $\alpha_{OX}$ is the point-to-point spectral slope between 2 keV and 2500 \AA \ in the restframe. The X-ray luminosities for the non-X-ray detected variable sources are upper limits, as indicated by the blue arrows.}
			\label{fig:LXLO}
		\end{figure}

\begin{figure}
			\centering
			\includegraphics[width=\columnwidth]{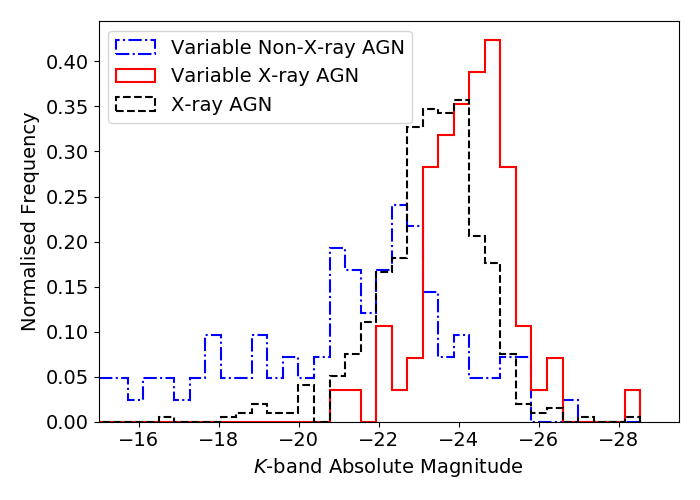}

			\caption{Normalised histogram of $K$-band absolute magnitudes showing that the 3 populations of AGN sample different luminosity ranges.}
			\label{fig:kbandhist}
		\end{figure}

		In addition, examining the X-ray to optical luminosity ratios of both the variable and X-ray AGN populations (Figure \ref{fig:LXLO}) shows that the variability method selects AGN that have lower luminosities (also shown in Figure \ref{fig:kbandhist}), and there is evidence that these variable AGN are systematically X-ray quiet as they have higher $\alpha_{OX}$ values.
		
		This evidence suggests that the AGN selection method presented here finds AGN that are systematically more X-ray quiet than the Chandra AGN, and therefore probes a low mass population that the deep Chandra data cannot detect. This results in a set of AGN that would be missed by the most common AGN selection technique in deep surveys.

    \subsection{IRAC AGN}
        \label{sec:irac}
		
\begin{figure}
			\centering
			\includegraphics[width=\columnwidth]{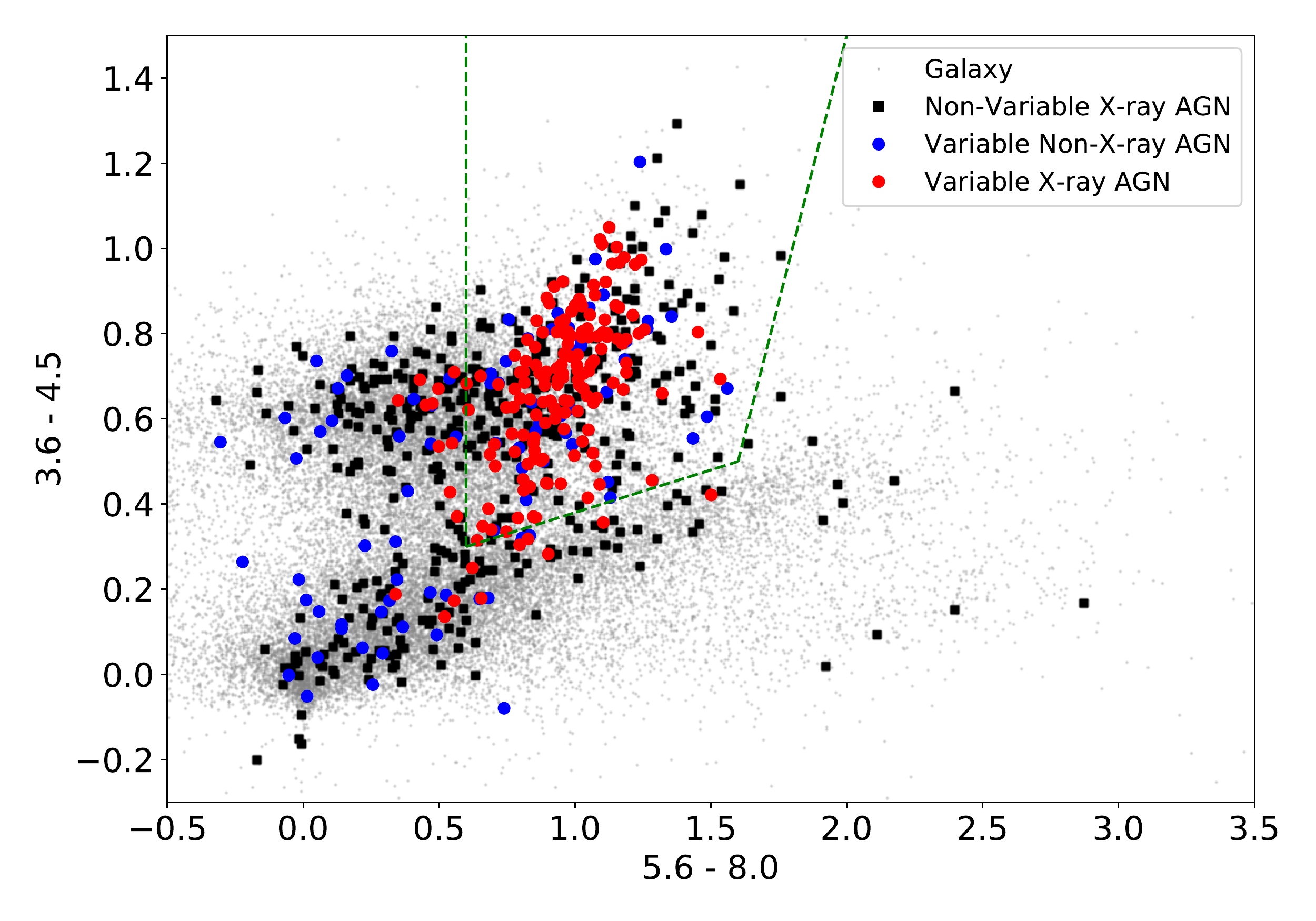}

			\caption{Stern AGN selection plot of mid-infrared colours. AGN are known to sit within the wedge indicated.}
			\label{fig:stern}
		\end{figure}

		The "Stern Wedge" \citep{Stern2005} is another frequently used method of identifying AGN, in this case through their IRAC colours. It uses the four IRAC bands to define a wedge within which sources are deemed AGN; the selection boundaries are shown by the green dashed lines in Figure \ref{fig:stern}. While there is significant contamination in this sample from star-forming galaxies at high and low redshifts \citep[for more details, see][]{Donley2008}, it is nonetheless useful to examine any overlap between the techniques. In total, there are 6417 sources in the UDS that meet the Stern AGN criteria: 188 of theses are also variable AGN. 
			
		When comparing the Stern and variability selected samples, it should be noted that a source must have detections in all four IRAC bands to be selected by the Stern method. Only 249 of the variable AGN meet that criterion, and of these 76\% are also Stern AGN.
						
\begin{figure}
			\centering
			\includegraphics[width=\columnwidth]{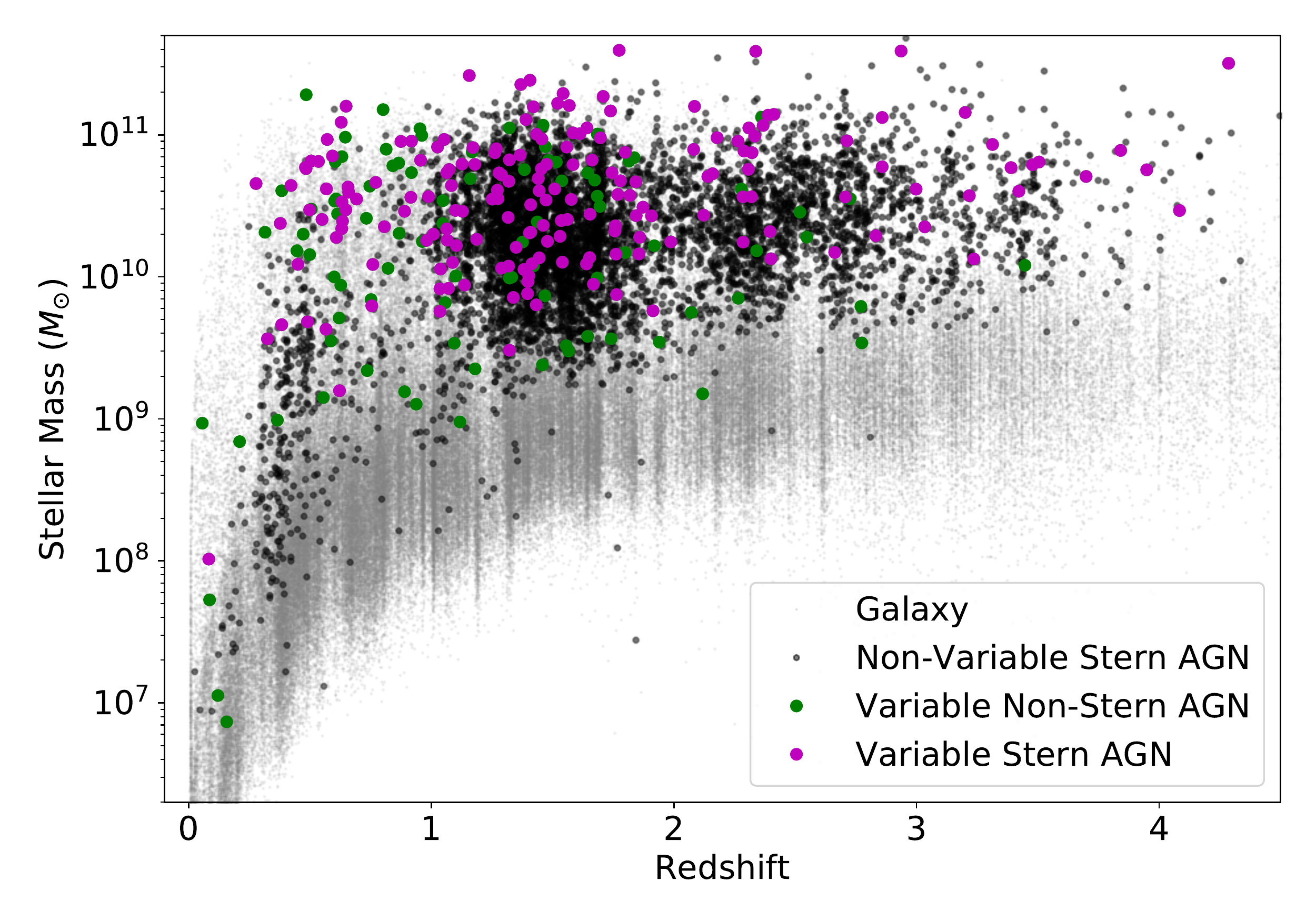}

			\caption{Stellar mass distribution with redshift showing where the Stern and variable AGN populations sit within the parameter space. Only those variables with detections in all 4 IRAC bands are shown here.}
			\label{fig:mstarstern}
		\end{figure}

		Figure \ref{fig:mstarstern} shows the same plot as Figure \ref{fig:mstarxray} with the Stern and variable AGN that have full IRAC detections highlighted. While the difference between those that are in both samples and those that are only seen in the variable sample is not as strong here, this is only because the low mass variable AGN that were seen in Figure \ref{fig:mstarxray} are not found in the IRAC catalogue. Therefore, the variability study probes a low mass region that is also missed by the Stern selection method.
			
    \subsection{Catalogue Comparison}
    
        The above comparisons have shown that the variability analysis seems to find AGN across a wider range in stellar mass than either X-ray or IRAC selection methods. There are a total of 185 variables in the full sample that are not detected by either of the other selection methods, and these typically have lower stellar masses and luminosities. 
        
		While the Stern method selects the most AGN, it is also the most contaminated sample. This contamination makes comparisons between the variable and Stern AGN samples redundant. Therefore, when comparing the physical properties of the different AGN samples in Section \ref{sec:properties}, we focus on the comparison with X-ray selected AGN.
		
\section{Characterising AGN}
    \label{sec:properties}

    \subsection{X-ray Detected and Non-X-ray Detected Variable Sources}
	    \label{sec:xraynonxray}
	
\begin{figure}
			\centering
			\includegraphics[width=\columnwidth]{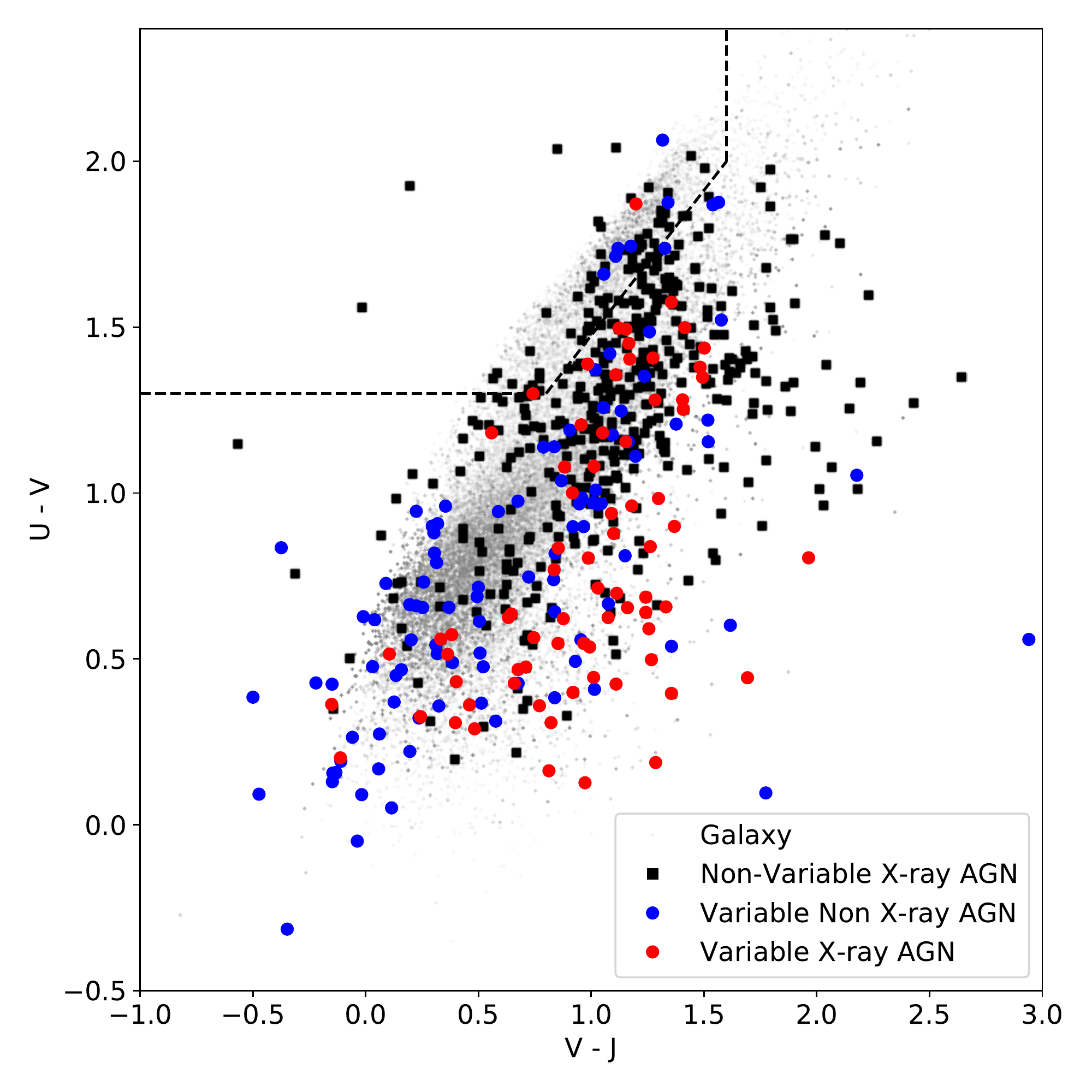}

			\caption{A rest-frame UVJ colour-colour diagram, used to differentiate quiescent and star-forming galaxies. Quiescent galaxies are found in the top left of the diagram, as defined by the dashed line, while star-forming galaxies and quasars are found in the remaining regions. The plot demonstrates that the majority of the variable AGN have rest-frame optical colours consistent with star-forming galaxy hosts.}
			\label{fig:uvjdiagram}
		\end{figure}

		As mentioned in Section \ref{sec:xray}, the variable non-X-ray detected sources appear to be a different class of AGN from the variable X-ray detected sources. When plotted on a rest-frame UVJ colour-colour diagram (Figure \ref{fig:uvjdiagram}), which is commonly used to differentiate quiescent and star-forming galaxies, the variable X-ray detected sources are generally bluer than ordinary galaxies in U-V, consistent with some AGN contribution in UV light, while the variable non-X-ray detected sources do not show this excess UV light in general, which might be due to a more obscured central AGN or a low-luminosity AGN (LLAGN).
		
\begin{figure}
			\centering
			\includegraphics[width=\columnwidth]{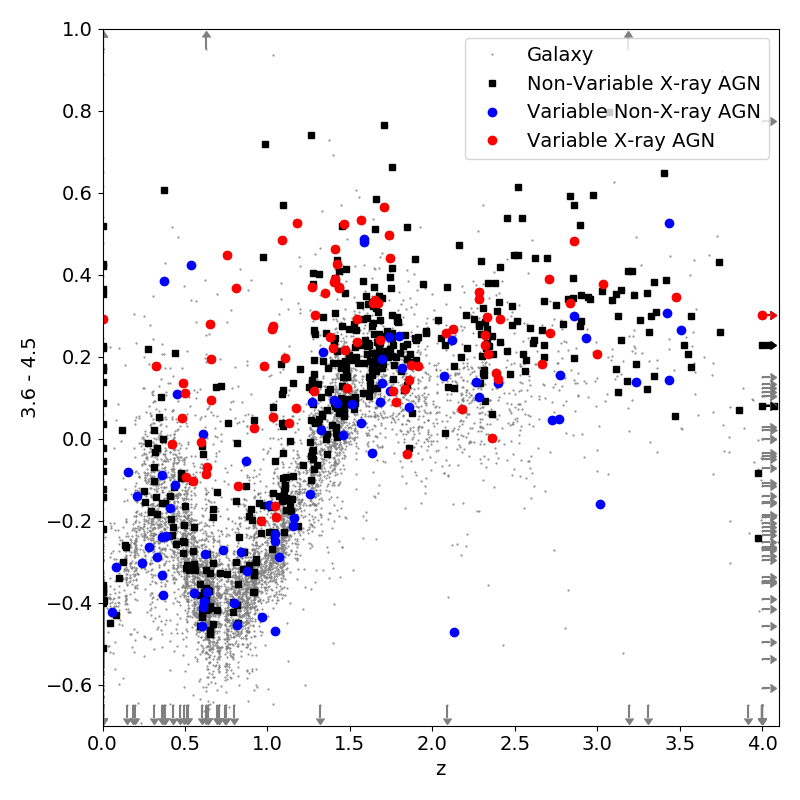}

			\caption{3.6 - 4.5 micron colour as a function of redshift. AGN usually sit above the galaxy locus in this space.}
			\label{fig:iracz}
		\end{figure}

		Looking into the IRAC colours of the populations beyond simply using them for selection can also provide insight into these sources. The variable non-X-ray detected sources are once again seen scattered through the galaxy locus in both Figures \ref{fig:stern} and \ref{fig:iracz}, while the variable X-ray detected sources are slightly offset. This adds further weight to the idea that these populations are dominated by different types of AGN.
		
		While we have spectroscopic redshifts for a a significant fraction of the variable sources (71 in total within the Chandra region, 56 X-ray detected and 15 non-X-ray detected), we only have access to 20 spectra for the variable sources; 12 of these are for the X-ray detected sources and 8 for the non-X-ray detected sources. The sources for these redshifts and spectra were discussed in Section \ref{sec:data}.
		
\begin{figure*}
			\centering
			\subfloat{\includegraphics[width=\linewidth]{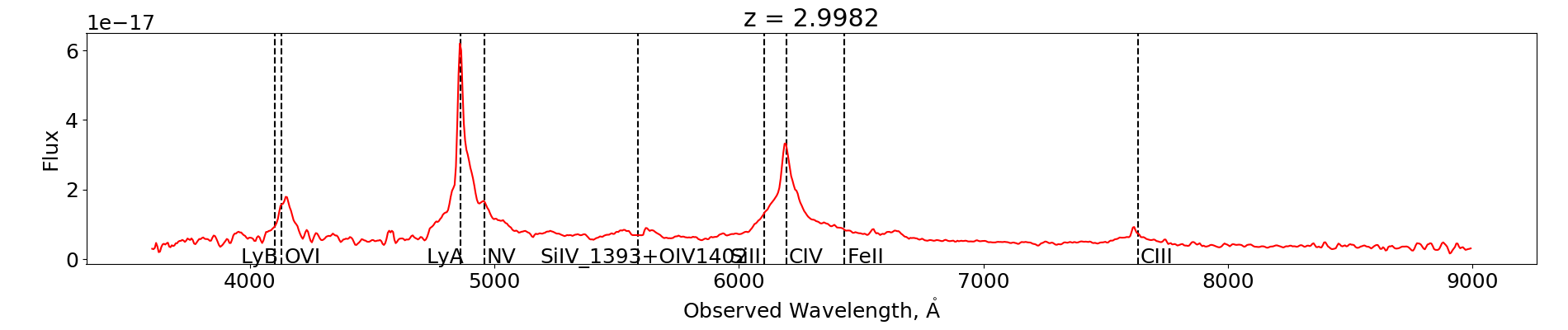}}

			\subfloat{\includegraphics[width=\linewidth]{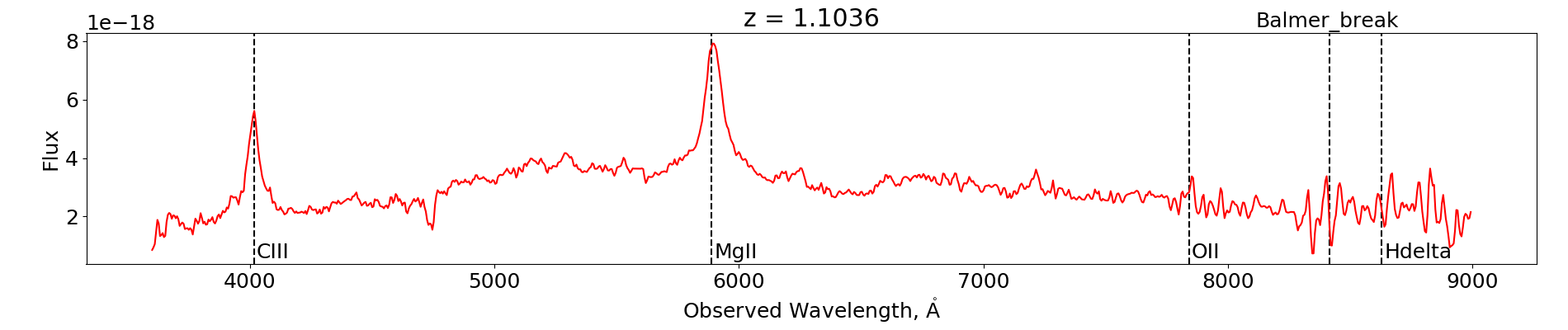}}

			\caption{Spectra of two X-ray detected variable AGN. Both spectra are VIMOS spectra from UDSz.}
			\label{fig:xspec}
		\end{figure*}

\begin{figure*}
			\centering
			\subfloat{\includegraphics[width=\linewidth]{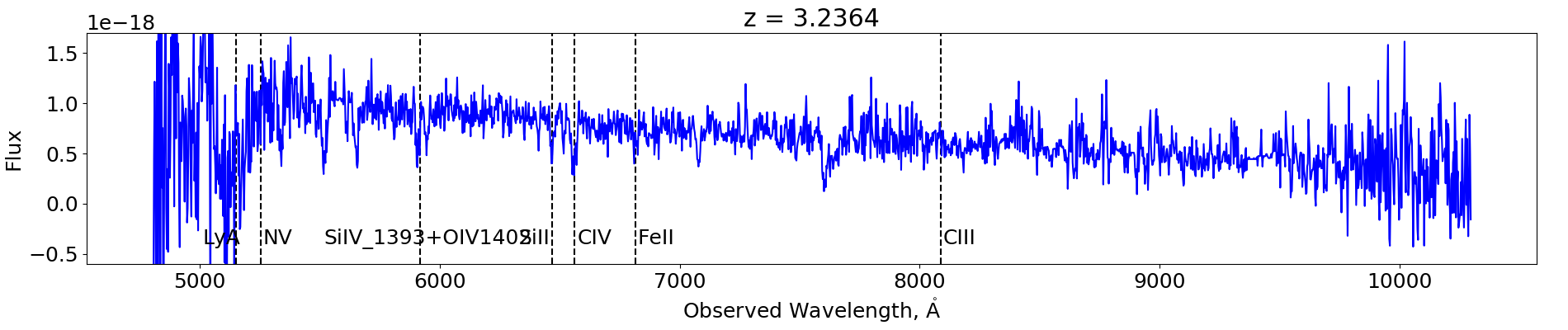}}
			
			\subfloat{\includegraphics[width=\linewidth]{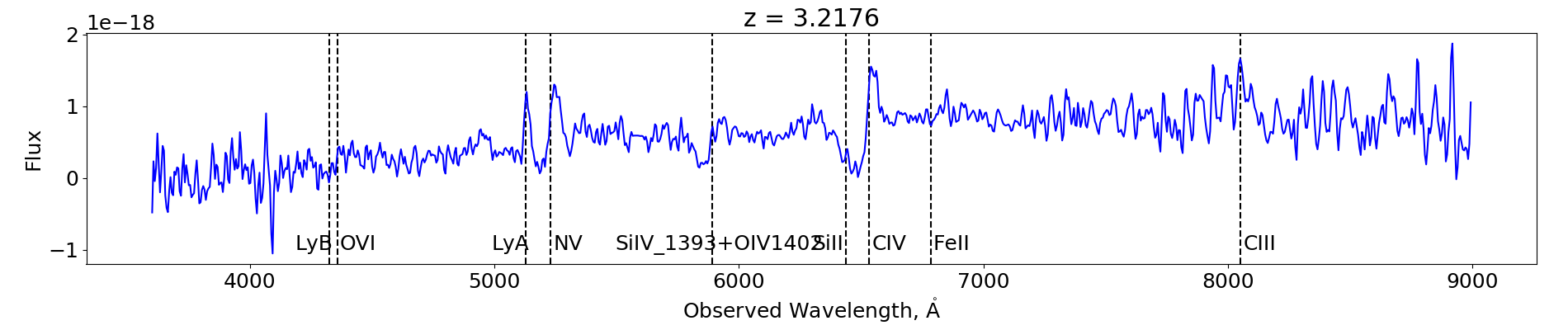}}
			
			\caption{Spectra of two variable AGN that are not detected in X-rays. The top figure shows a VLT VIMOS spectrum from VANDELS of a normal $z\sim3$ star-forming galaxy, while the bottom figure shows a VIMOS spectrum from UDSz that reveals a candidate BAL quasar.}
			\label{fig:nonxspec}
		\end{figure*}

\begin{figure}
			\centering
			\subfloat{\includegraphics[width=\linewidth]{mag_173520_notitle.png}}

			\subfloat{\includegraphics[width=\linewidth]{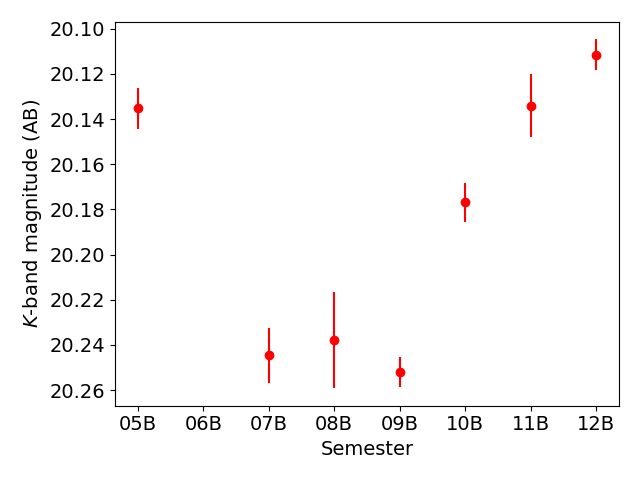}}

			\caption{The $K$-band light curves for the variable X-ray detected AGN whose spectra are shown in \ref{fig:xspec}}
			\label{fig:xlc}
		\end{figure}

\begin{figure}
			\centering
			\subfloat{\includegraphics[width=\linewidth]{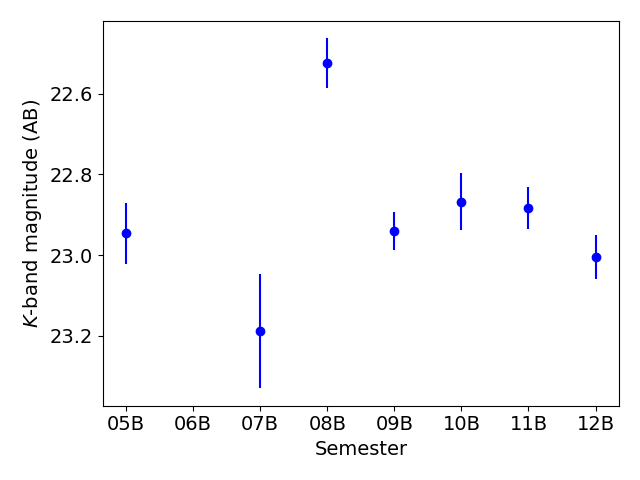}}

			\subfloat{\includegraphics[width=\linewidth]{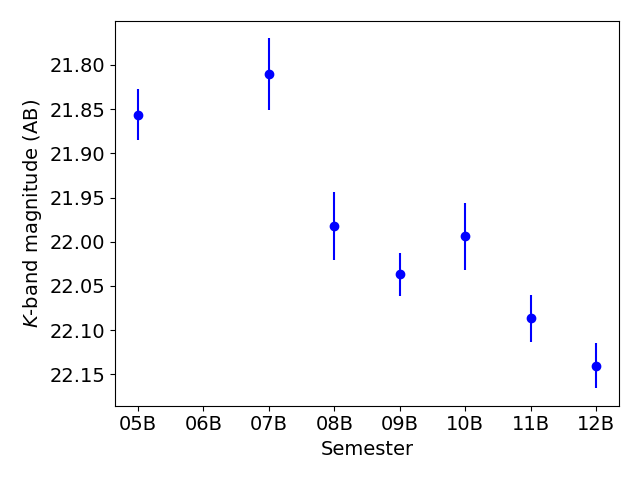}}

			\caption{The $K$-band light curves for the variable non-X-ray detected AGN whose spectra are shown in \ref{fig:nonxspec}}
			\label{fig:nonxlc}
		\end{figure}

		From examining the limited spectra available, we find that the spectra of variable non-X-ray detected sources show features consistent with a different AGN population to those of the variable X-ray detected sources. The variable X-ray detected sources are generally found to have broad line AGN spectra (e.g. Figure \ref{fig:xspec}), while the  spectra of variable non-X-ray sources are typically dominated by the light of the host galaxy (e.g. top of Figure \ref{fig:nonxspec}). Among the variable non-X-ray AGN we also identify a rare Broad Absorption Line (BAL) quasar \citep{Weymann1981}, showing the characteristic blue-shifted absorption in the CIV line (bottom of Figure \ref{fig:nonxspec}), which suggests that NIR variability could be effective at finding rare and interesting types of AGN. The light curves for those objects whose spectra are shown in Figures \ref{fig:xspec} and \ref{fig:nonxspec} are shown in Figures \ref{fig:xlc} and \ref{fig:nonxlc} respectively.
		
		The hypothesis that the variable non-X-ray detected sources are typically LLAGN is given further weight by Figures \ref{fig:LXLO} and \ref{fig:kbandhist}. These figures highlight that the variables not detected in X-rays have lower luminosities on average in the X-ray, optical and NIR regimes than the variables that are detected in X-rays. In addition, Figure \ref{fig:LXLO} shows that the variable non-X-ray sources tend to have higher $\alpha_{OX}$ values, which suggests that they are systematically more X-ray quiet than the variable X-ray detected AGN population. This leads to the conclusion that long-term NIR variability studies can highlight populations of LLAGN that are missed by other selection techniques.
		

	\subsection{NIR-variable and non-variable sources detected in X-rays}
	
	While about half of the variable sources are detected in X-rays, there are a large number of X-ray sources in the UDS that do not have significant variability in the NIR. Figure \ref{fig:chisquared} shows that the majority of the X-ray sources ($\sim 86\%$) sit within the main distribution of non-varying sources. This raises the question: what makes the X-ray sources that do have variable light curves different?
	
	From looking at the same colour plots as mentioned in Section \ref{sec:xraynonxray} (Figures \ref{fig:stern}, \ref{fig:uvjdiagram}, \ref{fig:iracz}), we can see that the NIR-non-variable X-ray sources (black squares) sit separately from the NIR-variable X-ray sources (red circles). Notably, the X-ray NIR-variables in Figure \ref{fig:stern} are mostly within the Stern wedge, but the NIR-non-variable X-ray sources are spread throughout the galaxy locus. The X-ray sources that are found in the variable sample, tend to be high luminosity sources (Figures \ref{fig:LXLO} and \ref{fig:kbandhist}) whose IRAC and optical colours appear to be dominated by the AGN, while the non-variable sources seem to be dominated by light from the host galaxy. While it may seem contradictory that the lower luminosity X-ray AGN are less likely to be found in this variability study, given previous works have found that lower luminosity AGN have higher variability amplitudes \citep[e.g.][]{MacLeod2010, Kozowski2016, Vagnetti2016}, it is likely that signal-to-noise provides an explanation for these trends. The lower luminosity AGN will be fainter and are more likely to be dominated by the host galaxy in the $K$-band, and therefore the variability from the nucleus will be harder to detect despite the larger amplitudes.

\section{A new population of AGN?}
	\label{sec:newpop}
	
	The new population of LLAGN identified by this variability analysis are relatively faint, therefore there may be concerns that these variations are simply noise, or that the variability indicates a different class of object (e.g. supernovae). In this section we address these potential concerns.
	
	\subsection{Noise?}
		
		While the light curves of low flux objects in the survey are noisier, the self-calibrated uncertainties applied to the data in the analysis (Section \ref{sec:selection}) mean that any increase in noise at low fluxes is accounted for in the uncertainty.  In addition, the threshold for the variability selection was chosen such that only 6 false positives are expected across the whole sample, therefore the entire contamination from false positives is less than 2\%.
	
	\subsection{Supernovae?}
	
\begin{figure}
			\centering
			\includegraphics[width=\linewidth]{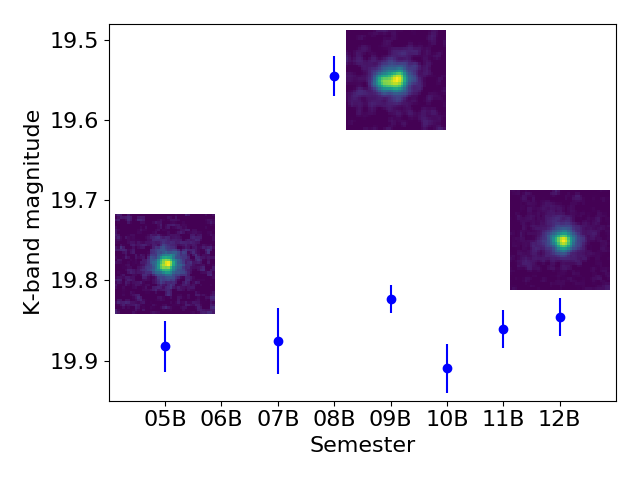}

			\caption{Light curve for a potential supernova within the variable sample. The inlaid images show the source before, during and after the flare, which can be seen as a separate object to the left of the galaxy.}
			\label{fig:sn}
		\end{figure}

		One other source of variability is supernovae. As the timescale for supernova events is less than one year, we can identify them by finding sources that only vary in a single epoch. To find these, a secondary analysis was run where we removed the most deviant point of the light curve (i.e. the furthest from the mean) and then recalculated $\chi^{2}$.
		We identified those sources that had a $<99.9\%$ chance of being truly variable (i.e. $\chi^{2} < 15.09$) after the removal, and checked these by eye to see if there was truly one flux value that was very different to the rest of the light curve.
		Fifty sources that had one significantly different epoch were identified, including one definite SN event at $z\sim 2$ (Figure \ref{fig:sn}). Removing these objects does not significantly change the mass and luminosity distributions and therefore we conclude that these light curves with anomalous epochs are not driving the trend for variability selected AGN to be found at lower masses and luminosities.
	
	\subsection{Low-Luminosity AGN?}
	
	    As we have ruled out the population being down to noise or supernova contamination, the most plausible explanation is that they are LLAGN. This explanation is reinforced by looking at the luminosities of the sources in X-rays, optical and the NIR (Figures \ref{fig:LXLO} and \ref{fig:kbandhist}). These figures clearly show that the variable non-X-ray detected sources have systematically lower luminosities in all three regimes.
	    
	    In addition, LLAGN are expected to have higher amplitudes of variability than high luminosity AGN. Studies in the X-ray regime \citep[e.g.][]{Lawrence1993, Almaini2000, Vagnetti2016} and in the UV/optical \citep[e.g.][]{MacLeod2010, Kozowski2016} show that the amplitude of variability is anti-correlated with the luminosity of the AGN. LLAGN have also been found using long-term variability selection methods in the optical and X-ray regimes \citep[e.g.][]{Sanchez-Saez2019, Young2012}, therefore it is perhaps not surprising that highly variable LLAGN are found among the low-mass host galaxies in this survey.

\section{Summary}
	 
    This paper has presented a new method of AGN selection using only the NIR light curves of objects in the UDS. This method involves applying a $\chi^{2}$ analysis to the light curves of non-stellar objects with detections in all 7 epochs. A selection threshold of $\chi^{2}=30$ was chosen in order to minimise false detections while maximising the number of variables identified. This gave a sample of 393 variable AGN across the UDS field. 

    A comparison to other selection methods showed that only a subset of the variability selected AGN are found in the X-ray (41\%) and Stern (48\%) selected AGN samples. Within the region of the field imaged with Chandra, those X-ray AGN that are variable tend to be in high mass, high luminosity hosts whose optical and IR light is dominated by the AGN. 
    In contrast, the variable AGN that are not detected in Chandra have lower masses and luminosities, and are dominated by the host galaxy. 
    
    We have accounted for any potential contamination of the sample of variable AGN due to noise or supernova events. In preparing the data, we removed any variations due to PSF changes between epochs by convolving the images to match the PSF of the worst image, and incorporated any contribution from noise within the self-calibrated photometric uncertainties. Light curves containing the pattern of variability expected in supernovae were identified through the secondary $\chi^{2}$ analysis, where the most deviant point was removed from each light curve. When these potential supernovae were removed from the sample, the mass and luminosity distributions of the variable, non-X-ray detected AGN remained lower than those of the X-ray AGN.
    
    Our work builds on studies that select AGN through their optical variability \citep[e.g.][]{Sanchez-Saez2019, Pouliasis2019, DeCicco2019}, but offers the unique ability to study rest frame optical variability at high redshifts ($z > 2$), and potentially dust-obscured AGN that may be missed by optical selection. Further follow up of the AGN identified in this study will allow a more detailed comparison.
    
    In summary, we find that selecting AGN through just their NIR variability finds a population of low-mass, low-luminosity, X-ray quiet AGN that would be missed by conventional selection techniques, even deep Chandra imaging, as the variability selection probes a wider range of masses and luminosities. Studying this population further will allow us to develop our understanding of the structure of AGN across this wide range of luminosities. We have also found that NIR variability can help us discover peculiar AGN, such as the BAL quasar shown in Figure \ref{fig:nonxspec}.
    
    Future work on this sample should include examining the light curves on shorter timescales, and looking at correlations between the $K$-band variability and the $J$ and $H$-band light curves that are also available for all objects in the UDS. These studies would allow us to quantify the structure of the variations, helping us understand the underlying processes behind AGN variability.

\section*{Acknowledgements}

EE is supported by a United Kingdom Science and Technology Facilities Council (STFC) studentship.
We extend our gratitude to the staff at UKIRT for their tireless efforts in ensuring the success of the UDS project. We also wish to recognise and acknowledge the very significant cultural role and reverence that the summit of Mauna Kea has within the indigenous Hawaiian community. We were most fortunate to have the opportunity to conduct observations from this mountain. This work is based in part on observations from ESO telescopes at the Paranal Observatory (programmes 094.A-0410, 180.A-0776 and 194.A-2003).




\bibliographystyle{mnras}
\bibliography{variability}




%
%


\bsp	
\label{lastpage}
\end{document}